\documentclass[12pt, a4paper]{article}
\usepackage{amsmath,amsfonts}
\usepackage{hhline}
\usepackage{indentfirst}
\usepackage{caption3}
\usepackage[pdfborder={0 0 0}]{hyperref}
\usepackage{tikz}
\usetikzlibrary{positioning}
\makeatletter
\renewcommand{\@oddhead}{Romanov V. F. (\href{mailto:romvf@mail.ru}{romvf@mail.ru})\hfil \thepage}
\makeatother

\begin{document}
\thispagestyle{empty}
\begin{center}
{\bf DISCORDANT COMPACT LOGIC-ARITHMETIC STRUCTURES IN~DISCRETE OPTIMIZATION PROBLEMS}
\end{center}
\begin{center}
{\bf Romanov V. F. (\href{mailto:romvf@mail.ru}{romvf@mail.ru})}
\end{center}

\begin{center}
{ Vladimir state university}
\end{center}

\begin{abstract}
In sphere of research of discrete optimization algorithms efficiency the important place occupies a method of polynomial reducibility of some problems to others with use of special purpose components. In this paper a novel method of compact representation for sets of binary sequences in the form of ``compact triplets structures" (CTS) and ``compact couples structures" (CCS) is stated, supposing both logic and arithmetic interpretation of data. It is shown that any non-empty CTS in dual interpretation represents some unique Boolean formula in 3-CNF and the tabular CTS contains all satisfyig sets of the formula as concatenations of the triplets chosen from the neighbouring tiers. In general, any 3-CNF formula is transformed by decomposition to a system of discordant CTS's, each being associated with an individual permutation of variables constructed by a polynomial algorithm. As a result the problem of the formula satisfiability is reduced to the following one: ascertain the fact of existence (or absence) of a ``joint satisfying set" (JSS) for all discordant structures, based on the different permutations. Further transformation of each CTS to CCS is used; correctness of preservation of the allowed sets is reached by simple algorithmic restrictions on triplets concatenation. Then the procedure of ``inverting of the same name columns" in the various structures is entered for the purpose of reducing the problem of JSS revealing to elementary detection of $n$-tuples of zeros in the CCS system. The formula is synthesized, being on the structure a variation of 2-CNF, associated with the calculation procedure realizing adaptation of the polynomial algorithm of constraints distribution (well-known in the optimization theory) to the efficient resolving Boolean formula coded by means of discordant compact structures.

\end{abstract}

\textbf{1. Introduction. Tabular formulas}

A large number of discrete optimization problems are combinatorial and certain are intractable. Analysis and classification of these problems often involve reducibility methods based on models using special constructive components. By this reason, new research results on models, properties, computational techniques, and algorithms for some selected intractable problems often assume generalization.

In this paper a non-orthodox method of compact represeentation for sets of binary sequences in the form of {\it compact triplets structures (CTS)} and {\it compact couples structures (CCS)} is suggested, supposing both logic and arithmetic interpretation of data. The suitable illustration of application of these structures  is a unique combinatorial model for the classic 3-Satisfiability problem(3-SAT) [1 - 3].

The problem statement: for  given  $m$  elementary disjunctions $C_1, C_2, \,\dots,\, C_m$, each containing exactly 3 literals, referring to Boolean variables $x_1, x_2, \,\dots,\, x_n$, determine, whether the formula $$F=C_1\wedge C_2\wedge \cdots \wedge C_m$$  is satisfiable or unsatisfiable.

We will use for the formula presented in the conjunctive normal form (CNF) a specific recording mode---the form of a table ({\it a tabular formula}), containing $n$ columns, noted by names of the variables, and  $m$ lines, each presenting the term $C_i$ by 0-1 sequence: 0  written in column  $j$  and line  $i$  marks the occurrence of $x_j$   in the term $C_i$ without negation, 1---with the sign of negation.
So, the tabular representation of the formula
$$
F=(\neg{x}_1 \vee x_2 \vee \neg{x}_4)\wedge
(x_2 \vee x_3 \vee \neg{x}_5)\wedge
(\neg{x}_3 \vee \neg{x}_4 \vee x_5)
$$ is as follows:

$$
\begin{array}{ccccccc}
&&F\\
x_1& x_2& x_3& x_4& x_5\\[3pt]
1&0&&1&\\
&0&0&&1&\\
&&1&1&0&.
\end{array}
$$

It is obvious that if 0 and 1 denote truth values: {\it false} and {\it true}, respectively (regardless of the denotation taken for the tabular formula\footnote{Actually we use dual interpretation of values 0 and 1 depending on a context.}), $F = 1$ {\bf for those and only for those sets of truth values, which do not contain any line from the tabular formula as a subset.}

\medskip
\textbf{2. Structures of compact triplets}

We consider for beginning a tabular formula consisting of terms in which three literals form {\it compact triplets (CT)}, that is  $\langle l_j$,~ $l_{j+1}$,~ $l_{j+2}\rangle$ sequences, where $l_j \in \{ x_j, \ \neg{x}_j\}$, $1\le j\le (n-2)$. We name such a formula a {\it CT formula {\rm (or} CTF}). The idea of CTF resolving is to transform the CTF to a structure of compact triplets (abbreviations: a {\it CT structure {\rm or} CTS}). The elements of the tabular CTS are lines, viz.\ {\it compact triplets} of variables' values located at  $n-2$  {\it tiers.} The tiers include variables numbered as 1,~2,~3; 2,~3,~4; \dots ; $n-2,~n-1,~n.$  Any CTS is composed of the triplets that are absent in the corresponding CTF, at each tier respectively. Generally, each tier contains a maximum of 8 binary lines. The final step of the CTS construction is a {\it clearing} procedure: removal from the tiers of {\it non-compatible lines}, i.e. the lines which cannot be adjoined to at least one line of each adjacent tier on condition that two values of variables written in succession coincide. The remaining lines are {\it compatible} and form sequences of length $n$  by means of {\it adjoining} operation (based on coincidence described above) applied to the pairs of lines from the tiers 1---2, 2---3, $\cdots,$ ($n-3$)---($n-2$). It is obvious that the CTF---CTS transformation is polynomial in terms of the algorithm complexity.

If at least one tier of the CTS turns out to be empty, the whole structure is declared an empty set of lines (or {\it an empty structure}) and the formula $F$ is declared a contradiction.
The CTS containing $n-2$ tiers can be formed if and only if $F$ is satisfiable. In fact, of a total of $2^n$ sequences of length $n$ all those and only those have been removed that include, as a subsequence, at least one line of the table representing formula $F$. Hence the CTS contains as the sequences of length $n$ all sets of truth values at which  $F$  is true (called {\it satisfying sets}).  Thus, {\it the very fact of existence of the CTS including  $n-2$  tiers means  that  $F$  is a satisfiable formula.}

Let us say that a structure of compact triplets is complete if each its tier contains 8 possible combinations of the binary values; such a structure represents the totality of $2^n$ satisfying sets ({\it SS}).

For example, the transformation of CTF  $F_1$  leads to the CTS  $Z$; the intermediate structure $Z^*$  still contains non-compatible lines (marked ``{\bf --}").

$$
\begin{array}{ccccccccccccccccccccccc}
&& F_1 &&&&& && Z^* &&&&& &&& Z &&\\[3pt]
x_1&x_2&x_3&x_4&x_5&&&x_1&x_2&x_3&x_4&x_5&&&&x_1&x_2&x_3&x_4&x_5\\[3pt]
0&0&0&&&&& 0&1&0&&&-&&& 0&1&1&&\\
0&0&1&&&&& 0&1&1&&&&&& 1&0&0&&\\
1&0&1&&&&& 1&0&0&&&&&& &1&1&0&\\
1&1&1&&&&& 1&1&0&&&-&&& &0&0&1&\\
&0&0&0&&&& &0&0&1&&&&& &&1&0&1\\
&1&0&0&&&& &0&1&0&&-&&& &&0&1&1\\
&1&0&1&&&& &0&1&1&&-&&& &&&&\\
&1&1&1&&&& &1&1&0&&&&& &&&&\\
&&0&1&0&&& &&0&0&0&-&&& &&&&\\
&&1&0&0&&& &&0&0&1&-&&& &&&&\\
&&1&1&1&&& &&0&1&1&&&& &&&&\\
&&&&&&& &&1&0&1&&&& &&&&\\
&&&&&&& &&1&1&0&-&&& &&&&\\
\end{array}
$$

Thus, analyzing $Z$, we fix two satisfying sets for $F$: 01101, 10011.

\medskip

{\it A substructure} $S'$ of the CTS $S$ is CTS composed of the subsets of the lines that are compatible at the adjacent tiers of $S$ (the notation: $S' \subseteq S$). Note that a substructure, that is not empty, consists of $n-2$ tiers like any CTS.
   We define an {\it elementary} CTS as a CTS containing only one line at each tier. An elementary CTS corresponds to a single set of truth values and, accordingly, to a single SS. 

Let $S_1,\ S_2,\ \ldots ,\ S_q$ be the system of CT structures based on {\it different permutations} of the variables (referred to as {\it discordant structures}). We define a $q$-ary operation of {\it unification} for the system as a special kind of a concretization of some variables values in accordance with the next rules of joint transformation of the structures:

\hangindent=1.1cm
1)  If some variable $x_j\equiv 0$ or $x_j\equiv 1$ in at least one CTS $S_p$ ($1\le j\le n$, $1\le p\le q$), then all the lines containing inverse value of this variable have to be removed from all CT structures.

\hangindent=1.1cm
2)  If two variables $x_j$ and $x_r$ appear together (in any order) in compact triplets inside two or more CT structures, then the values combinations for these variables must be the same in all such structures. All the lines that are in contradiction with this constraint have to be removed from these structures.

\hangindent=1.1cm
3)  The clearing procedure accompanies each event of lines removal from the CT structures.

Herewith, variants appear when the unified system turns out to be an empty set:
\begin{itemize}
\item a certain tier in at least one CTS turns out to be empty;
\item there exists a ``conflict" of constant
values for a certain variable $x_j$ in at least two CT structures, i.e. $x_j\equiv 0$ in one structure and  $x_j\equiv 1$
in another structure;
\item at least one CTS is originally empty (a trivial case).
\end{itemize}

Thus, the unified CT structures can be either empty or non-empty only simultaneously.

\par\medskip

\textbf{3. Formula decomposition}

In general case it is necessary for 3-SAT problem resolution to decompose the initial formula $F$  using the operation: $F=F_1\wedge F_2\wedge  \cdots \wedge F_k$,  where $F_r$ , ${r=1,\ 2,\, \ldots,\ k,}\ k\le m$, is the formula suitable for CTF presentation based on the individual permutation of variables ${P_r=\langle x_{r_1},\ x_{r_2}, \ldots,\ x_{r_n}\rangle}$.

The decomposition requires a polynomial procedure which consists of following points:
\begin{itemize}
\item grouping the lines of  $F$  with identical numbers of three
	non-empty columns;
\item	putting three non-empty columns including the symbols of the variables in each of $k$ obtained groups ($k$ matrices) into the places 1, 2, 3 with shifting the other columns; it causes fixation of  $k$ permutations of the variables as bases for  $k$  CTF.
\end{itemize}

So, the final  $k$  matrices are the ordinary CT formulas. Note that empty tiers are permitted in CTF, in contrast to CTS.

The described procedure comes to $k$-tuple survey of the lines of $nm$-matrix, hence the estimation of the complexity of a decomposition algorithm is $O(mnk)$. The suitable permutations are obtained by forming and not by enumeration; that results in elimination of exponential computation complexity.

The modernized algorithms can be based on different methods of assembling CTF out of matrices consisting of the first three columns of the CT formulas obtained by the previous algorithm; these columns are considered as tiers in lesser quantity of CTF.

Clearly, the parameter $k$ satisfies the condition  $\lceil w/(n-2)\rceil \le k \le m$, where  $w$  is the number of groups containing terms (the elementary disjunctions) with identical variables. For an ``ideal" formula  $F$,  $k = 1$; the extreme value $k = m$  relates to forming a separate permutation for each term of the initial formula. Note that we put aside possible methods of minimizing  $k$ as a non-principal point of the model realization.

Then we transform each CTF $F_r$ to CTS  $S_r$. Now the problem is reduced to the following one: ascertain the fact of existence (or absence) of {\it joint satisfying sets} (abbreviations: {\it JS sets} or {\it JSS}) for the system of discordant CT structures
$S_1,\ S_2,\, \dots, \ S_k$.  It is necessary to solve this new problem without a searching through the sets, coded in the CT structures, in order to avoid procedures of exponential complexity.

In order to illustrate theoretical aspects of the model realization (without restriction of the general analysis) we use, as an example, the initial tabular formula $F$ shown in Table 1.

The decomposition of $F$ was carried out with the use of assembling the tiers obtained by the procedure stated above. The resulting CT formulas based on three variable's permutations are presented in Table 2.

 Finally, CTF $\to$ CTS transformation described at Section 2 leads to the three CT structures:  $S_1$, $S_2$ and $S_3$ (Table 3).

\begin{table}[p]
\begin{tabular}{|c|c|c|c|c|c|c|c|c|c|c|c|c|c|c|c|c|c|c|c|c|c|c|c|c|c|}
\multicolumn{26}{c}{\bf Table 1. Initial formula $F$}\\
\multicolumn{26}{c}{}\\
\cline{1-8}
\cline{10-17}
\cline{19-26}
$a$&$b$&$c$&$d$&$e$&$f$&$g$&$h$&&$a$&$b$&$c$&$d$&$e$&$f$&$g$&$h$&&$a$&$b$&$c$&$d$&$e$&$f$&$g$&$h$\\
\hhline{|=|=|=|=|=|=|=|=|~|=|=|=|=|=|=|=|=|~|=|=|=|=|=|=|=|=|}
0&0&0&&&&&&&0&&1&&&&&0&&&&1&1&0&&&\\
\cline{1-8}                                                        \cline{10-17}
\cline{19-26}
0&1&0&&&&&&&1&&0&&&&&1&&&&1&0&&0&&\\
\cline{1-8}                                                        \cline{10-17}
\cline{19-26}
0&1&1&&&&&&&1&&1&&&&&1&&&&0&0&&0&&\\
\cline{1-8}                                                        \cline{10-17}
\cline{19-26}
1&0&0&&&&&&&&1&&&&&0&0&&&&1&&0&&&1\\
\cline{1-8}                                                        \cline{10-17}
\cline{19-26}
1&1&1&&&&&&&&0&&&&&1&0&&&&1&&0&&&0\\
\cline{1-8}                                                        \cline{10-17}
\cline{19-26}
0&1&&&0&&&&&&0&&&&&0&1&&&&&0&1&0&&\\
\cline{1-8}                                                        \cline{10-17}
\cline{19-26}
1&&&&1&0&&&&&1&&&&&0&1&&&&&1&0&1&&\\
\cline{1-8}                                                        \cline{10-17}
\cline{19-26}
1&&&&0&1&&&&&1&&&&&1&1&&&&&&0&0&0&\\
\cline{1-8}                                                        \cline{10-17}
\cline{19-26}
0&&0&&&0&&&&&0&&&0&&0&&&&&&&1&0&0&\\
\cline{1-8}                                                        \cline{10-17}
\cline{19-26}
0&&0&&&1&&&&&1&&&1&&1&&&&&&&1&1&1&\\
\cline{1-8}                                                        \cline{10-17}
\cline{19-26}
1&&1&&&0&&&&&0&&&0&&1&&&&&&&&0&1&0\\
\cline{1-8}                                                        \cline{10-17}
\cline{19-26}
1&&0&&&1&&&&&1&&&1&&&1&&&&&&&1&0&1\\
\cline{1-8}                                                        \cline{10-17}
\cline{19-26}
1&&&0&&0&&&&&0&&&0&&&0&&&&1&0&0&&&\\
\cline{1-8}                                                        \cline{10-17}
\cline{19-26}
0&&&0&&1&&&&&1&&&0&&0&&&0&&&1&&1&&\\
\cline{1-8}                                                        \cline{10-17}
\cline{19-26}
1&&&0&&1&&&&&&0&0&0&&&\\
\cline{1-8}                                                        \cline{10-17}

\end{tabular}
\end{table}

\begin{table}[p]
               \begin{tabular}{|c|c|c|c|c|c|c|c|c|c|c|c|c|c|c|c|c|c|c|c|c|c|c|c|c|c|}
\multicolumn{26}{c}{\bf Table 2. CT formulas}\\
\multicolumn{26}{c}{$F_1$\hspace{5cm}$F_2$\hspace{5cm}$F_3$}\\
\cline{1-8}
\cline{10-17}
\cline{19-26}
$a$&$b$&$c$&$d$&$e$&$f$&$g$&$h$&&$h$&$g$&$b$&$e$&$a$&$f$&$c$&$d$&&
$d$&$f$&$a$&$c$&$h$&$e$&$b$&$g$\\
\hhline{|=|=|=|=|=|=|=|=|~|=|=|=|=|=|=|=|=|~|=|=|=|=|=|=|=|=|}
0&0&0&&&&&&&0&0&1&&&&&&&0&0&1&&&&&\\
\cline{1-8}                                                        \cline{10-17}
\cline{19-26}
0&1&0&&&&&&&0&1&0&&&&&&&0&1&0&&&&&\\
\cline{1-8}                                                        \cline{10-17}
\cline{19-26}
0&1&1&&&&&&&1&0&0&&&&&&&0&1&1&&&&&\\
\cline{1-8}                                                        \cline{10-17}
\cline{19-26}
1&0&0&&&&&&&1&0&1&&&&&&&1&1&0&&&&&\\
\cline{1-8}                                                        \cline{10-17}
\cline{19-26}
1&1&1&&&&&&&1&1&1&&&&&&&&0&0&0&&&&\\
\cline{1-8}                                                        \cline{10-17}
\cline{19-26}
&&0&0&0&&&&&&0&0&0&&&&&&&0&1&1&&&&\\
\cline{1-8}                                                       \cline{10-17}
\cline{19-26}
&&1&0&0&&&&&&1&1&1&&&&&&&&0&1&0&&&\\
\cline{1-8}                                                        \cline{10-17}
\cline{19-26}
&&1&1&0&&&&&&1&0&0&&&&&&&&1&0&1&&&\\
\cline{1-8}                                                        \cline{10-17}
\cline{19-26}
&&&0&1&0&&&&&&1&0&0&&&&&&&1&1&1&&&\\
\cline{1-8}                                                        \cline{10-17}
\cline{19-26}
&&&1&0&1&&&&&&&1&1&0&&&&&&&1&1&0&&\\
\cline{1-8}                                                        \cline{10-17}
\cline{19-26}
&&&&0&0&0&&&&&&0&1&1&&&&&&&1&0&0&&\\
\cline{1-8}                                                        \cline{10-17}
\cline{19-26}
&&&&1&0&0&&&&&&&0&0&0&&&&&&&0&0&0&\\
\cline{1-8}                                                        \cline{10-17}
\cline{19-26}
&&&&1&1&1&&&&&&&0&1&0&&&&&&&1&1&1&\\
\cline{1-8}                                                        \cline{10-17}
\cline{19-26}
&&&&&0&1&0&&&&&&1&0&1&&&&&&&&0&1&0\\
\cline{1-8}                                                        \cline{10-17}
\cline{19-26}
&&&&&1&0&1&&&&&&1&1&0&&&&&&&&1&1&1\\
\cline{1-8}
\cline{10-17}
\cline{19-26}
\multicolumn{8}{}{}&&&&&&&0&1&0\\

\cline{10-17}

\multicolumn{8}{}{}&&&&&&&0&0&0\\

\cline{10-17}

\end{tabular}
\end{table}

\bigskip

\begin{table}[p]
\begin{tabular}{|c|c|c|c|c|c|c|c|c|c|c|c|c|c|c|c|c|c|c|c|c|c|c|c|c|c|}
\multicolumn{26}{c}{\bf Table 3. CT structures}\\
\multicolumn{26}{c}{$S_1$\hspace{5cm}$S_2$\hspace{5cm}$S_3$}\\
\cline{1-8}
\cline{10-17}
\cline{19-26}
$a$&$b$&$c$&$d$&$e$&$f$&$g$&$h$&&$h$&$g$&$b$&$e$&$a$&$f$&$c$&$d$&&
$d$&$f$&$a$&$c$&$h$&$e$&$b$&$g$\\
\hhline{|=|=|=|=|=|=|=|=|~|=|=|=|=|=|=|=|=|~|=|=|=|=|=|=|=|=|}
0&0&1&&&&&& &0&0&0&&&&&& &0&0&0&&&&&\\
\cline{1-8}                                                        \cline{10-17}
\cline{19-26}
1&0&1&&&&&& &0&1&1&&&&&& &1&0&0&&&&&\\
\cline{1-8}                                                        \cline{10-17}
\cline{19-26}
1&1&0&&&&&& &1&1&0&&&&&& &1&0&1&&&&&\\
\cline{1-8}                                                        \cline{10-17}
\cline{19-26}
&0&1&0&&&&& &&0&0&1&&&&& &1&1&1&&&&&\\
\cline{1-8}                                                        \cline{10-17}
\cline{19-26}
&0&1&1&&&&& &&1&0&1&&&&& &&0&0&1&&&&\\
\cline{1-8}                                                        \cline{10-17}
\cline{19-26}
&1&0&0&&&&& &&1&1&0&&&&& &&0&1&0&&&&\\
\cline{1-8}                                                        \cline{10-17}
\cline{19-26}
&1&0&1&&&&& &&&0&1&0&&&& &&1&1&0&&&&\\
\cline{1-8}                                                        \cline{10-17}
\cline{19-26}
&&0&0&1&&&& &&&0&1&1&&&& &&1&1&1&&&&\\
\cline{1-8}                                                        \cline{10-17}
\cline{19-26}
&&0&1&0&&&& &&&1&0&1&&&& &&&0&1&1&&&\\
\cline{1-8}                                                        \cline{10-17}
\cline{19-26}
&&0&1&1&&&& &&&&1&0&0&&& &&&1&0&0&&&\\
\cline{1-8}                                                        \cline{10-17}
\cline{19-26}
&&1&0&1&&&& &&&&1&0&1&&& &&&1&1&0&&&\\
\cline{1-8}                                                        \cline{10-17}
\cline{19-26}
&&1&1&1&&&& &&&&0&1&0&&& &&&&1&1&1&&\\
\cline{1-8}                                                        \cline{10-17}
\cline{19-26}
&&&0&1&1&&& &&&&1&1&1&&& &&&&0&0&0&&\\
\cline{1-8}                                                        \cline{10-17}
\cline{19-26}
&&&1&0&0&&& &&&&&0&0&1&& &&&&0&0&1&&\\
\cline{1-8}                                                        \cline{10-17}
\cline{19-26}
&&&1&1&0&&& &&&&&0&1&1&& &&&&1&0&1&&\\
\cline{1-8}                                                        \cline{10-17}
\cline{19-26}
&&&1&1&1&&& &&&&&1&0&0&& &&&&&1&1&0&\\
\cline{1-8}                                                        \cline{10-17}
\cline{19-26}
&&&&0&0&1&& &&&&&1&1&1&& &&&&&0&0&1&\\
\cline{1-8}                                                        \cline{10-17}
\cline{19-26}
&&&&1&0&1&& &&&&&&0&1&1& &&&&&0&1&0&\\
\cline{1-8}                                                        \cline{10-17}
\cline{19-26}
&&&&1&1&0&& &&&&&&0&0&1& &&&&&0&1&1&\\
\cline{1-8}                                                        \cline{10-17}
\cline{19-26}
&&&&&0&1&1& &&&&&&1&1&0& &&&&&&1&0&1\\
\cline{1-8}                                                        \cline{10-17}
\cline{19-26}
&&&&&1&0&0& &&&&&&1&1&1& &&&&&&0&1&1\\
\cline{1-8}                                                        \cline{10-17}
\cline{19-26}
\multicolumn{17}{}{}   &&&&&&&1&0&0\\
\cline{19-26}
\multicolumn{17}{}{}   &&&&&&&1&1&0\\
\cline{19-26}
\end{tabular}
\end{table}

\begin{table}[tbp]
\centering
\begin{tabular}{|c|c|c|c|c|c|c|c|c|c|c|c|c|c|c|c|c|}
\multicolumn{17}{c}{\bf Table 4. Unified CT structures \,$S_1$ and  $S_2$}\\
\multicolumn{17}{c}{$S_1$\hspace{5cm}$S_2$}\\
\cline{1-8}
\cline{10-17}
$a$&$b$&$c$&$d$&$e$&$f$&$g$&$h$&&$h$&$g$&$b$&$e$&$a$&$f$&$c$&$d$\\
\hhline{|=|=|=|=|=|=|=|=|~|=|=|=|=|=|=|=|=|}
0&0&1&&&&&& &0&0&0&&&&&\\
\cline{1-8}                                                        \cline{10-17}
1&0&1&&&&&& &1&1&0&&&&&\\
\cline{1-8}                                                        \cline{10-17}
&0&1&0&&&&& &&0&0&1&&&&\\
\cline{1-8}                                                        \cline{10-17}
&0&1&1&&&&& &&1&0&1&&&&\\
\cline{1-8}                                                        \cline{10-17}
&&1&0&1&&&& &&&0&1&0&&&\\
\cline{1-8}                                                        \cline{10-17}
&&1&1&1&&&& &&&0&1&1&&&\\
\cline{1-8}                                                        \cline{10-17}
&&&0&1&1&&& &&&&1&0&0&&\\
\cline{1-8}                                                        \cline{10-17}
&&&1&1&0&&& &&&&1&0&1&&\\
\cline{1-8}                                                        \cline{10-17}
&&&1&1&1&&& &&&&1&1&1&&\\
\cline{1-8}                                                        \cline{10-17}
&&&&1&0&1&& &&&&&0&0&1&\\
\cline{1-8}                                                        \cline{10-17}
&&&&1&1&0&& &&&&&0&1&1&\\
\cline{1-8}                                                        \cline{10-17}
&&&&&0&1&1& &&&&&1&1&1&\\
\cline{1-8}                                                        \cline{10-17}
&&&&&1&0&0& &&&&&&0&1&1\\
\cline{1-8}                                                        \cline{10-17}
\multicolumn{8}{}{}    &&&&&&&1&1&0\\
                                                        \cline{10-17}
\multicolumn{8}{}{}    &&&&&&&1&1&1\\
                                                        \cline{10-17}
\end{tabular}
\end{table}

\textbf {4. Solution of JSS existence problem for two CT structures}

\smallskip
The resolution of 3-SAT problem for the formula reduced to two CT structures is a clue to the solution of the general problem. Let  $S_1$  and  $S_2$  be the two CT structures based on different permutations of variables (we use the structures from Table 3). The primary stage of CTS processing consists in unification operation for $S_1$ and $S_2$. This operation simplifies the CTS-operands by removing some of the lines that do not belong to JS sets, but preserves JS sets (if there exist any) in accordance with the operation rules. By this reason, we do not change notation for the unified CT structures (Table 4).

Let $S_1$ be a {\it basic structure}; we fix for it the initial numeration of variables: $x_1, \,x_2$, \,\ldots ,\,~$x_n$
($a, \,b$, \,\ldots ,\ $h$, in the presented example). 

The determination of JSS for $S_1$ and $S_2$
denotes satisfiability of the formula $F'$ presented by the subset of the lines in Table 1 (the lines that served for forming  $S_1$  and  $S_2$  before the unification of these structures).

Let's consider a trivial special case: the triplets 000 are present at all tiers of the compared CT structures that may be easily discovered algorithmically. It is obvious that the difficulties caused by the search with different permutations are thus excluded: JSS represents a nil-set of length $n$. The offered approach to the problem decision is based on this fact 
\footnote{In this work we refer to the distinguished set  00...0  formed of $n$  zeros as the nil-set.}. 

We will enter into consideration an inversion operator for the column elements of CTS (briefly:  {\it column inversion operator}) and a binary vector of {\it columns inversion control (CIC)}. We place the CIC vectors over the headings of each CTS, keeping the conformity of components in all permutations. The component ``0" in the CIC vector means the column preservation, the component ``1" means the column inversion.
 
The described means allow to modify CTS. The algorithm of full searching through $2^n$  CIC vectors guarantees detection of all nil-JSS for modified structures or ascertaining the fact of JSS absence, moreover, is thus formed the information about JSS for the initial structures. 

For example, CIC vectors: 10101100 for $S_1$ and, accordingly, 00011110 for $S_2$, generate modified structures $S^*_1$ and $S^*_2$  (Table 5). Now the triplets 000 are present at all tiers of the compared structures.

REMARK 4.1. The top lines of the two structures in the Table 5 are the same CIC vector 
applied to two permutations of the variables.
 
REMARK 4.2. Each new CIC vector generated at search is applied to inverting of columns of each initial CTS. Detection in all structures of elementary CTS coinciding with a nil-set at some step of search fixes a JSS for the initial CTS system, thus the JSS itself on construction coincides with the CIC vector.  

We will dwell upon last statement using for the illustration Tables 4 and 5. CTS S1 and S2 con-tain JSS 10101100 in terms of initial numbering of variables, but it is not apparent because of per-mutations difference in two structures. 

The suitable CIC vector generated at searching removes symbols 1 in the specified JSS, trans-forming it into a nil-set which is easily discovered in various structures. It is possible (from the very definition of a CIC vector) only in that case when symbols 1 in CIC vector are located on the same places as in the stated JSS. 

It will be hereinafter evidently displayed dualism in interpretation of equally designated data, noted at the very beginning of the article. Depending on a context and the purpose of application symbols 0 and 1 are considered as logic values {\it false} and {\it true} or as binary arithmetic values, in particular, components of binary sequences and vectors. Such interpretation makes a conceptual basis of the offered research.

\begin{table}[h]
\centering
\begin{tabular}{|c|c|c|c|c|c|c|c|c|c|c|c|c|c|c|c|c|}
\multicolumn{17}{c}{\bf Table 5. Modified CT structures \,$S^*_1$ and $S^*_2$ }\\
\multicolumn{17}{c}{$S^*_1$\hspace{5cm}$S^*_2$}\\
\cline{1-8}
\cline{10-17}
$1$&$0$&$1$&$0$&$1$&$1$&$0$&$0$&&$0$&$0$&$0$&$1$&$1$&$1$&$1$&$0$\\
\cline{1-8}                                                        \cline{10-17}
$a$&$b$&$c$&$d$&$e$&$f$&$g$&$h$&&$h$&$g$&$b$&$e$&$a$&$f$&$c$&$d$\\
\hhline{|=|=|=|=|=|=|=|=|~|=|=|=|=|=|=|=|=|}
1&0&0&&&&&& &0&0&0&&&&&\\
\cline{1-8}                                                        \cline{10-17}
0&0&0&&&&&& &1&1&0&&&&&\\
\cline{1-8}                                                        \cline{10-17}
&0&0&0&&&&& &&0&0&0&&&&\\
\cline{1-8}                                                        \cline{10-17}
&0&0&1&&&&& &&1&0&0&&&&\\
\cline{1-8}                                                        \cline{10-17}
&&0&0&0&&&& &&&0&0&1&&&\\
\cline{1-8}                                                        \cline{10-17}
&&0&1&0&&&& &&&0&0&0&&&\\
\cline{1-8}                                                        \cline{10-17}
&&&0&0&0&&& &&&&0&1&1&&\\
\cline{1-8}                                                        \cline{10-17}
&&&1&0&1&&& &&&&0&1&0&&\\
\cline{1-8}                                                        \cline{10-17}
&&&1&0&0&&& &&&&0&0&0&&\\
\cline{1-8}                                                        \cline{10-17}
&&&&0&1&1&& &&&&&1&1&0&\\
\cline{1-8}                                                        \cline{10-17}
&&&&0&0&0&& &&&&&1&0&0&\\
\cline{1-8}                                                        \cline{10-17}
&&&&&1&1&1& &&&&&0&0&0&\\
\cline{1-8}                                                        \cline{10-17}
&&&&&0&0&0& &&&&&&1&0&1\\
\cline{1-8}                                                        \cline{10-17}
\multicolumn{8}{}{}    &&&&&&&0&0&0\\
                                                        \cline{10-17}
\multicolumn{8}{}{}    &&&&&&&0&0&1\\
                                                        \cline{10-17}
\end{tabular}
\end{table}

The full searching through CIC vectors leads to discovering of five JSS in terms of two permutations:

$$
\begin{array}{cccccccc}
a&b&c&d&e&f&g&h\\[5pt]
0&0&1&0&1&1&0&0\\
0&0&1&1&1&1&0&0\\
1&0&1&0&1&1&0&0\\
1&0&1&1&1&1&0&0\\
0&0&1&1&1&0&1&1\\
\end{array}
$$

{

\medskip

$$
\begin{array}{cccccccc}
h&g&b&e&a&f&c&d\\[5pt]
0&0&0&1&0&1&1&0\\
0&0&0&1&0&1&1&1\\
0&0&0&1&1&1&1&0\\
0&0&0&1&1&1&1&1\\
1&1&0&1&0&0&1&1\\ 
\end{array}
$$

\noindent and to conclusion that formula  $F'$  is satisfiable. 

\newpage

The procedure of JSS determination for the input formula $F$ uses the unified CT structures  $S_1$, $S_2$ and $S_3$ presented in Table 6 (based on the structures from Table 3). For the formula  $F$  two JSS are discovered: 00111011 and 10111100 at initial numbering of variables ($a, \,b$, \,\ldots ,\ $h$).

\begin{table}[h]
\begin{tabular}{|c|c|c|c|c|c|c|c|c|c|c|c|c|c|c|c|c|c|c|c|c|c|c|c|c|c|}
\multicolumn{26}{c}{\bf Table 6. Unified CT structures \,$S_1$, $S_2$,  $S_3$}\\
\multicolumn{26}{c}{$S_1$\hspace{5cm}$S_2$\hspace{5cm}$S_3$}\\
\cline{1-8}
\cline{10-17}
\cline{19-26}
$a$&$b$&$c$&$d$&$e$&$f$&$g$&$h$&&$h$&$g$&$b$&$e$&$a$&$f$&$c$&$d$&&
$d$&$f$&$a$&$c$&$h$&$e$&$b$&$g$\\
\hhline{|=|=|=|=|=|=|=|=|~|=|=|=|=|=|=|=|=|~|=|=|=|=|=|=|=|=|}
0&0&1&&&&&& &0&0&0&&&&&& &1&0&0&&&&&\\
\cline{1-8}
\cline{10-17}
\cline{19-26}
1&0&1&&&&&& &1&1&0&&&&&& &1&1&1&&&&&\\
\cline{1-8}
\cline{10-17}
\cline{19-26}
&0&1&1&&&&& &&0&0&1&&&&& &&0&0&1&&&&\\
\cline{1-8}
\cline{10-17}
\cline{19-26}
&&1&1&1&&&& &&1&0&1&&&&& &&1&1&1&&&&\\
\cline{1-8}
\cline{10-17}
\cline{19-26}
&&&1&1&0&&& &&&0&1&0&&&& &&&0&1&1&&&\\
\cline{1-8}
\cline{10-17}
\cline{19-26}
&&&1&1&1&&& &&&0&1&1&&&& &&&1&1&0&&&\\
\cline{1-8}
\cline{10-17}
\cline{19-26}
&&&&1&0&1&& &&&&1&0&0&&& &&&&1&1&1&&\\
\cline{1-8}
\cline{10-17}
\cline{19-26}
&&&&1&1&0&& &&&&1&1&1&&& &&&&1&0&1&&\\
\cline{1-8}
\cline{10-17}
\cline{19-26}
&&&&&0&1&1& &&&&&0&0&1&& &&&&&1&1&0&\\
\cline{1-8}
\cline{10-17}
\cline{19-26}
&&&&&1&0&0& &&&&&1&1&1&& &&&&&0&1&0&\\
\cline{1-8}
\cline{10-17}
\cline{19-26}
\multicolumn{8}{}{}    &&&&&&&0&1&1& &&&&&&1&0&1\\

\cline{10-17}
\cline{19-26}
\multicolumn{8}{}{}    &&&&&&&1&1&1& &&&&&&1&0&0\\
                                                        \cline{10-17}
\cline{19-26}
\end{tabular}
\end{table}

Let's notice that the stated theory do not undergo basic changes in a case of large values $n$ and $k$.

Further we introduce a method of JSS revealing for the set of compact structures without use of exponential procedures.

\par\medskip

\textbf{5. Structures of compact couples}

By analogy with CTS structures, we will enter into consideration for any 3-CNF formula a {\it system of compact couples structures (CCS)} which principle of construction follows the above-stated concept for compact triplets structures with some natural differences. 

The elements of the tabular CCS are lines, viz. {\it compact couples} of variables' values located at  $n-1$  tiers. Generally, each tier contains a maximum of 4 binary lines (00, 01, 10, 11). The adjoining operation for the couples situated at adjacent tiers is based on coincidence of values only by one variable. A general totality of binary sequences of length $n$  formed by all possible concatenations of  4  lines at each tear corresponds to $2^n$ sets coded by CCS (as well as by CTS).

Any CTS can be transformed to CCS by splitting of each compact triplet  $x_j x_{j+1}x_{j+2}$  to two couples. with one common element: $x_j x_{j+1}$ and  $x_{j+1}x_{j+2}$ , ${j=1,\ 2,\, \ldots,\ n-2}$, and placing these couples, presented by the values, on adjaicent $jth$ and $(j+1)th$ tears. The lines at any tier are not duplicated.

In this particular example CCS $G_1$ and $G_2$, put in conformity with unified CTS $S_1$ and $S_2$ (Table 4), are shown in Table 7.

Notice that each CCS, received by transformation of CTS, on construction contains all binary sets coded in corresponding CTS, and, generally, still some new sets ({\it superfluous sets}). For example, $G_1$  includes the concatenation 00101011 that is absent in $S_1$. 

\begin{table}[tbp]
\centering
\begin{tabular}{|c|c|c|c|c|c|c|c|c|c|c|c|c|c|c|c|c|c|}
\multicolumn{17}{c}{\bf Table 7. Structures of compact couples \,$G_1$ and  $G_2$}\\
\multicolumn{17}{c}{$G_1$\hspace{5cm}$G_2$}\\
\cline{1-8}
\cline{10-17}
$a$&$b$&$c$&$d$&$e$&$f$&$g$&$h$&&$h$&$g$&$b$&$e$&$a$&$f$&$c$&$d$\\
\hhline{|=|=|=|=|=|=|=|=|~|=|=|=|=|=|=|=|=|}
0&0&&&&&&& &0&0&&&&&&\\
\cline{1-8}                                                        \cline{10-17}
1&0&&&&&&& &1&1&&&&&&\\
\cline{1-8}                                                        \cline{10-17}
&0&1&&&&&& &&0&0&&&&&\\
\cline{1-8}                                                        \cline{10-17}
&&1&0&&&&& &&1&0&&&&&\\
\cline{1-8}                                                        \cline{10-17}
&&1&1&&&&& &&&0&1&&&&\\
\cline{1-8}                                                        \cline{10-17}
v&&&0&1&&&&&&&&1&0&&&\\
\cline{1-8}                                                        \cline{10-17}       
&&&1&1&&&&&&&&1&1&&&\\                                
\cline{1-8}                                                        \cline{10-17}
v&&&&1&0&&&& &&&&0&0&&\\
\cline{1-8}                                                        \cline{10-17}
&&&&1&1&&&& &&&&0&1&&\\
\cline{1-8}                                                        \cline{10-17}
&&&&&0&1&& &&&&&1&1&&\\
\cline{1-8}                                                        \cline{10-17}
&&&&&1&0&&&w& &&&&0&1&\\
\cline{1-8}                                                        \cline{10-17}
&&&&&&1&1&& &&&&&1&1&\\
\cline{1-8}                                                        \cline{10-17}
&&&&&&0&0& &&&&&&&1&1\\
\cline{1-8}                                                        \cline{10-17}
\multicolumn{8}{}{}    &&w&&&&&&1&0\\
                                                        \cline{10-17}

\end{tabular}
\end{table}

The cause of appearing of superfluous sets is emergence in CCS at the structures transformation of new (in comparison with CTS) concatenations of the lines at the adjacent tiers that correspond to compact triplets nonexistent in CTS. In Table 7 the pairs of  the lines, which joint inclusion in JSS is inadmissible, are noted in the first columns of the structures by the labels with the same name: v and w, accordingly, ({\it twins labels}); indeed, the triplets $0\,1\,0$ ($d\,e\,f$) in $S_1$ and $0\,1\,0$ ($fc\,d\,$) in $S_2$ are absent. The simple algorithmic analysis of admissibility of the lines concatenations in CCS by their comparison with triplets in CTS which have served to its generation should be completed with arrangement of the labels mentioned above (necessarily before modification of the structures).

The existence of inadmissible pairs of lines itself is not an obstacle for JSS searching algorithm as any line with a label forms on construction concatenations with lines without the twin label, 
located on the adjacent tiers: these concatenations are those compact triplets which have generated compact couples. Therefore the twins labels play for the algorithm a role of "indicators" of an incorrect choice of the line pairs which should be avoided; theoretically, inevitability of such choice is a ``deadlock" (a kind of a contradiction).

This consideration applies to the general case of the structures transformation for large values $n$ and $k$.

\par
\medskip

\textbf{6. Polynomial procedure for decision of a problem of JSS existence}

The final stage of constructive approach to the problem decision is based on modification of CCS system for the purpose of display of a nil-JSS, with simultaneous building of vector $U$ (CIC vector) for the initial system, or stating the fact of JSS absence meaning that the initial CNF is unsatisfiable.
The logic formula

\begin{equation}
\bigwedge_{t=1}^{m_1} \neg (s_{i_t} \vee v_{j_t}),
\end{equation}

where: literals  $s_{i_t}$  and  $v_{j_t}$   designate  different variables of the problem or their inversions, 

$m_1$ is  the number of disjunctions formed with use of CCS system, $m_1 \le (n-1)k$,\\
serves a basis for description of the actions intended to built a binary nil-set.

The basic idea of the procedure operation is to form a new truth assignment for the variables of CCS so that some JSS (if there exists any) becomes apparent in form of a nil-set. 
At the beginning all variables are assigned in accordance with initial CC structures. Preliminary step is inverting of all ``constant 1 columns" accompanied by placing the values 1 into the corresponding components of vector $U$. Then a starting variable and its value are to be chosen arbitrary.

Certain steps of the algorithmic detailed elaboration of the procedure are {\it residue points}, at which a {\it residual problem} is stated and new starting variable is chosen which can be given either of two different values. Later steps in the algorithm may cause it to return to one of these residue points (it is a {\it  backtracking}). However, only the most recent starting variable can be backtracked over; all choices made earlier than the most recent one are permanent. 

It is necessary here to describe {\bf the scheme of distribution of zero values;} we use the starting variable for an explanation, though the scheme applies to any regular stage of CCS system processing until the next residue point is reached.

Let it be $x_{i_1 }= 0$  for any literal in (1). In CCS system all the couples, containing  $x_{i_1 }= 0$, have to be found (no matter, at the first or second place). Let at some tier (tiers) a variable  $x_j$  is present 
together with  $x_{i_1 }$  ($j \ne i_1$). As is shown in fig.\,1,  $x_{i_1 }$    can be present at corresponding tiers in two couples or in one: 
  
$$
\begin{array}{cccccccccc}

x_{i_1 }& x_j&& x_{i_1 }& x_j&&x_{i_1 }& x_j\\[3pt]
0&0&&0&0&&0&1\\
0&1&&&&&&\\[4pt]
&a)&&&b)&&&c)
\end{array}
$$

\begin{center}
Figure 1: Variants of the couple presence
\end{center}

In case $(a)$ the variable  $x_j$  should keep both its values, it may be specified later or remain with two possible values; that means possibility of the admissible decisions with alternative values of  $x_j$. In case $(b)$ $x_j=0$  is to be set; in case $(c)$   $x_j  = 0$ is to be set and vector  $U$  inverts columns  $x_j$  in all structures turning values 1 into 0. The described actions for   $x_j$   also concern the permutation  $x_j \, x_{i_1 }$.
 
If for the starting variable $x_{i_1 }$ value 1 is chosen then corresponding columns have to be inverted 
($\neg x_{i_1 }= 0$) and value 1 moves into the component of vector $U$; further the new variables forming couples with  $\neg x_{i_1 }$ have to be found et cetera (as at  $x_{i_1 } = 0$).

The logic consequence of the first step (use of zero value of a starting variable) is setting of zero values for some variables in couples; that is equivalent to the enumeration of brackets with zero literals $s_{i_t}$  and  $v_{j_t}$   in (1). These variables, in turn, continue under the offered scheme generation of new variables with zero values in modified CCS system etc. 

 Notice: there are no backtrackings in this part of procedure. The described setting of zero values is absolutely compulsory. 

The procedure purpose is a true value of the formula (1) (in arithmetic interpretation it is 1). The disjunctions in (1) are not preassigned originally but are formed in the course of the analysis of CCS tables. The formation of each couple $0\,0$  as a content of the brackets in (1) may be described by implication $(s_{i_t}=0) \Rightarrow (v_{j_t}=0)$ under the condition similar to cases $(b)$  and $(c)$ (fig. 1). 

\bf {The variants of continuation and completion of the procedure.}

\rm 1. A contradiction for constant values of some variable is revealed at least in two various CCS as the consequence of the used value settings. Then the algorithm backtracks to its most recent residue point, undoing the assignments it made since that choice, and renovates the decision with use of distribution scheme. If the processing leads to a contradiction again, or if the algorithm has already backtracked over the most recent residue point, then it aborts the search and reports that the input formula is unsatisfiable.

2. One of the alternative values of the starting variable initiates with use of distribution scheme  suitable setting of values for all variables, displaying a nil-set. This is a variant of the most effective realisation of the procedure caused by initial data.
 
3. For the current starting variable the process of the assignments has been expired, being  completed for a subset $ X_1   \subset X$ , and no contradiction has been fixed. For the set $X_2$ =$X\backslash X_1$  we have the problem which is not solved yet. It is a residue point. Next, the procedure chooses a new starting variable for the set  $X_2 \subset  X$  and continues application of the scheme of zero values distribution for variables $x_{i_t}\in X_2$ , with use of the modified CCS system, existing at the beginning of this step as the starting structures for a residual problem. The variables from the set  $X_1$  and the corresponding columns of the structures do not change any more; the problem for  $X_2$  is autonomous and determines the final result. 
 
The process of formation of new subsets    $X_3,\ X_4,\, \dots, \ X_q$  and, accordingly, new starting variables, may continue. The algorithm may use backtracking only for the most recent starting variable. The last residual problem for the subset $X_q$ determines the result of the input formula classification.

REMARK 6.1. The set  $X_q$ may include a subset  $X'_q$ consisting of the variables for which any of two values are suitable (fig.\,1,\,$a$) and may be arbitrary chosen to complete the assignments if the formula is classified as satisfiable.

So, we have described a principle of successive reduction of the main problem size. 

\medskip

\bf {Example 1.} \rm The decision of a problem for two CCS  $G_1$ and $G_2$ presented in Table 7}.

\smallskip

 We begin with $a = 0$. In Table 8 the result of the starting vector $U=00101000$ action is shown: the replacement of constant 1 columns  $c$  and  $e$  by constant 0 columns.

The analysis of the columns containing the lines with  $a = 0$  in the modified structures $G_1^*$ and $G_2^*$  leads to paying attention to couples $a\, b = 0\, 0$, and $e\, a = 0\, 0$, not involving any changes of the tables 
(because $b$ and $e$ are constants), and also to $a\, f = 0\, 0$ and $a\, f = 0 \,1$. Two last couples do not cause
   the  setting of $f$ (see fig.\,1,\,$a$)   and this variable at the given step remains unassigned.

So, at $ a = 0$  the action of the starting variable  $a$  comes to an end with preservation of all columns (besides $c$  and  $e$) in the modified structures. No contradiction emergence means the choice correctness: testing of the alternative value  $a = 1$  later on is not required.
The letter ``T" ({\it terminal})  in the tables marks the columns that are completely formed.

\begin{table}[tbp]
\centering
\begin{tabular}{|c|c|c|c|c|c|c|c|c|c|c|c|c|c|c|c|c|c|}
\multicolumn{17}{c}{\bf Table 8. The partition of the set $X$ }\\
\multicolumn{17}{c}{$G_1^*$\hspace{5cm}$G_2^*$}\\
\cline{1-8}
\cline{10-17}
$a$&$b$&$c$&$d$&$e$&$f$&$g$&$h$&&$h$&$g$&$b$&$e$&$a$&$f$&$c$&$d$\\
\hhline{|=|=|=|=|=|=|=|=|~|=|=|=|=|=|=|=|=|}
0&0&&&&&&& &0&0&&&&&&\\
\cline{1-8}                                                        \cline{10-17}
1&0&&&&&&& &1&1&&&&&&\\
\cline{1-8}                                                        \cline{10-17}
&0&0&&&&&& &&0&0&&&&&\\
\cline{1-8}                                                        \cline{10-17}
&&0&0&&&&& &&1&0&&&&&\\
\cline{1-8}                                                        \cline{10-17}
&&0&1&&&&& &&&0&0&&&&\\
\cline{1-8}                                                        \cline{10-17}
v&&&0&0&&&&&&&&0&0&&&\\
\cline{1-8}                                                        \cline{10-17}       
&&&1&0&&&&&&&&0&1&&&\\                                
\cline{1-8}                                                        \cline{10-17}
v&&&&0&0&&&& &&&&0&0&&\\
\cline{1-8}                                                        \cline{10-17}
&&&&0&1&&&& &&&&0&1&&\\
\cline{1-8}                                                        \cline{10-17}
&&&&&0&1&& &&&&&1&1&&\\
\cline{1-8}                                                        \cline{10-17}
&&&&&1&0&&&w& &&&&0&0&\\
\cline{1-8}                                                        \cline{10-17}
&&&&&&1&1&& &&&&&1&0&\\
\cline{1-8}                                                        \cline{10-17}
&&&&&&0&0& &&&&&&&0&1\\
\cline{1-8}                                                        \cline{10-17}
T&T&T&&T&&&& &w&&&&&&0&0\\

\cline{1-8}                                                        \cline{10-17}
                                                        
\multicolumn{8}{}{}    &&&&T&T&T&&T&\\
                                                        \cline{10-17}

 \end{tabular}
\end{table}

The procedure proceeds to a residual problem that deals with variables $d, f, g, h$.

Now we begin with  $f$ = 0 for this reduced problem (any choice of a new starting variable and its value is correct). 

$(f = 0) \Rightarrow (\neg g = 0) \Rightarrow (\neg h = 0)$  (see columns  $f , g, h$  in CCS $G_1^*$);

$(f = 0) \Rightarrow (\neg d = 0)$  (see columns $f, c, d$  in CCS $G_2^*$: at $f = 0$ the labels allow for couple  $c\, d$  only 
a penultimate line; the constant  $c = 0$  does not enter into a subset for the reduced problem and plays 
a passive transit role for the values transfer).
The modification of the structures on the basis of implications results is shown in Table 9. 

\begin{table}[tbp]
\centering
\begin{tabular}{|c|c|c|c|c|c|c|c|c|c|c|c|c|c|c|c|c|c|}
\multicolumn{17}{c}{\bf Table 9. The final modification of the structures} \\
\multicolumn{17}{c}{$G_1^{**}$\hspace{5cm}$G_2^{**}$}\\
\cline{1-8}
\cline{10-17}
$a$&$b$&$c$&$d$&$e$&$f$&$g$&$h$&&$h$&$g$&$b$&$e$&$a$&$f$&$c$&$d$\\
\hhline{|=|=|=|=|=|=|=|=|~|=|=|=|=|=|=|=|=|}
0&0&&&&&&& &1&1&&&&&&\\
\cline{1-8}                                                        \cline{10-17}
1&0&&&&&&& &0&0&&&&&&\\
\cline{1-8}                                                        \cline{10-17}
&0&0&&&&&& &&1&0&&&&&\\
\cline{1-8}                                                        \cline{10-17}
&&0&1&&&&& &&0&0&&&&&\\
\cline{1-8}                                                        \cline{10-17}
&&0&0&&&&& &&&0&0&&&&\\
\cline{1-8}                                                        \cline{10-17}
v&&&1&0&&&&&&&&0&0&&&\\
\cline{1-8}                                                        \cline{10-17}       
&&&0&0&&&&&&&&0&1&&&\\                                
\cline{1-8}                                                        \cline{10-17}
v&&&&0&0&&&& &&&&0&0&&\\
\cline{1-8}                                                        \cline{10-17}
&&&&0&1&&&& &&&&0&1&&\\
\cline{1-8}                                                        \cline{10-17}
&&&&&0&0&& &&&&&1&1&&\\
\cline{1-8}                                                        \cline{10-17}
&&&&&1&1&&&w& &&&&0&0&\\
\cline{1-8}                                                        \cline{10-17}
&&&&&&0&0&& &&&&&1&0&\\
\cline{1-8}                                                        \cline{10-17}
&&&&&&1&1& &&&&&&&0&0\\
\cline{1-8}                                                        \cline{10-17}
T&T&T&T&T&T&T&T& &w&&&&&&0&1\\

\cline{1-8}                                                        \cline{10-17}
                                                        
\multicolumn{8}{}{}    &&T&T&T&T&T&T&T&T\\
                                                        \cline{10-17}

 \end{tabular}
\end{table} 

The procedure of the assignment for all variables is completed. The final vector $U$ = 00111011 represents JSS for formula $F'$.

For completeness of a picture as a whole we consider the procedure action for the alternative value of the starting variable: $a = 1 \ (\neg a = 0)$. The simple analysis of Table 8 columns with inversion of the columns entitled by the variable $a$, generates implications:

$(\neg a = 0) \Rightarrow (\neg f = 0) \Rightarrow (g = 0) \Rightarrow (h = 0)$;
 
a current problem for $d$: $d$ = * (any of two values is admissible). So, vector $U$ = 101*1100 represents two joint satisfying sets.
 
Two more JSS are expressed by vector $U$ = 001*1100 (the starting values for the problem and 
a residual problem are presented by the first and sixth components:  $a = 0,  f = 1$).

\medskip
\bf {Example 2.} \rm The problem 3-SAT solving for the input formula $F$ .

\smallskip
Table 10 is built on the base of Table 6 for the unified CTS. The twins labels are present only in CTS $H_3$.
In order to present a visual inference, in the initial Table 10 the constant values $c = 1$ 
and $e = 1$ are inverted and, accordingly, the constant 1 columns are replaced by 0 columns 
($U$ = 00101000). The twins labels forbid the compact triplets  $a\,c\, h = 0\,1\,0$ and  $a\,c\, h = 1\,1\,1$ (but not $0\,0\,0$ and  $a\,c\, h = 1\,0\,1$ ), since the labels have been assigned before the column  $c$  inversion. 

The couples in Table 10 presented in thick typing explain the next implications that compose the effective conclusion with  $a = 0$:

 $(a = 0)\Rightarrow (f = 0)\Rightarrow (\neg d = 0)$;

 $(f=0)\Rightarrow (\neg g = 0) \Rightarrow (\neg h = 0).$ 

\begin{table}[tbp]
\begin{tabular}{|c|c|c|c|c|c|c|c|c|c|c|c|c|c|c|c|c|c|c|c|c|c|c|c|c|c|}
\multicolumn{26}{c}{\bf Table 10. Unified CCS \,$H_1$, $H_2$,  $H_3$}\\
\multicolumn{26}{c}{$H_1$\hspace{5cm}$H_2$\hspace{5cm}$H_3$}\\
\cline{1-8}
\cline{10-17}
\cline{19-26}
$a$&$b$&$c$&$d$&$e$&$f$&$g$&$h$&&$h$&$g$&$b$&$e$&$a$&$f$&$c$&$d$&&
$d$&$f$&$a$&$c$&$h$&$e$&$b$&$g$\\
\hhline{|=|=|=|=|=|=|=|=|~|=|=|=|=|=|=|=|=|~|=|=|=|=|=|=|=|=|}
0&0&&&&&&& &0&0&&&&&&& &\bf {1}&\bf {0}&&&&&&\\
\cline{1-8}
\cline{10-17}
\cline{19-26}
1&0&&&&&&& &1&1&&&&&&& &1&1&&&&&&\\
\cline{1-8}
\cline{10-17}
\cline{19-26}
&0&0&&&&&& &&0&0&&&&&& &&0&0&&&&&\\
\cline{1-8}
\cline{10-17}
\cline{19-26}
&&0&1&&&&& &&1&0&&&&&& &&1&1&&&&&\\
\cline{1-8}
\cline{10-17}
\cline{19-26}
&&&1&0&&&& &&&0&0&&&&& &v&&0&0&&&&\\
\cline{1-8}
\cline{10-17}
\cline{19-26}
&&&&0&0&&& &&&&0&0&&&& &w&&1&0&&&&\\
\cline{1-8}
\cline{10-17}
\cline{19-26}
&&&&0&1&&& &&&&0&1&&&& &w&&&0&1&&&\\
\cline{1-8}
\cline{10-17}
\cline{19-26}
&&&&&\bf {0}&\bf {1}& &&&&&&\bf {0}&\bf {0}& &&&v&&&0&0&&&\\  
\cline{1-8}
\cline{10-17}
\cline{19-26}
&&&&&1&0&& &&&&&1&1&&& &&&&&1&0&&\\
\cline{1-8}
\cline{10-17}
\cline{19-26}
&&&&&&\bf {1}&\bf {1}& &&&&&&0&0&& &&&&&0&0&&\\
\cline{1-8}
\cline{10-17}
\cline{19-26}
&&&&&&0&0&   &&&&&&1&0&& &&&&&&0&0&\\
\cline{1-8}
\cline{10-17}
\cline{19-26}
\multicolumn{8}{}{} &&&&&&&&0&1 &&&&&&&&0&1\\
\cline{10-17}
\cline{19-26}
\multicolumn{8}{}{}&&&&&&&&&   &&&&&&&&0&0\\
\cline{10-17}
\cline{19-26}                                              

\end{tabular}
\end{table}

Table 11 displays the results of zero distribution in the CCS system. Each tier in every CCS contains the couple $0\,0$ (thick typing), the concatenations of these couples represent the nil-JSS in the modified structures. The corresponding joint satisfying set for the initial structures coincides with the vector  $U$  = 00111011.

Naturally, the first appearance of JSS in the course of the structures modification is sufficient for stating of the input formula satisfiability; that in some cases implies the early end of the procedure.

\begin{table}[tbp]
\begin{tabular}{|c|c|c|c|c|c|c|c|c|c|c|c|c|c|c|c|c|c|c|c|c|c|c|c|c|c|}
\multicolumn{26}{c}{\bf Table 11. The transformed CCS \,$H_1^*$, $H_2^*$,  $H_3^*$}\\
\multicolumn{26}{c}{$H_1^*$\hspace{5cm}$H_2^*$\hspace{5cm}$H_3^*$}\\
\cline{1-8}
\cline{10-17}
\cline{19-26}
$a$&$b$&$c$&$d$&$e$&$f$&$g$&$h$&&$h$&$g$&$b$&$e$&$a$&$f$&$c$&$d$&&
$d$&$f$&$a$&$c$&$h$&$e$&$b$&$g$\\
\hhline{|=|=|=|=|=|=|=|=|~|=|=|=|=|=|=|=|=|~|=|=|=|=|=|=|=|=|}
\bf {0}&\bf {0}&&&&&&& &1&1&&&&&&& &\bf {0}&\bf {0}&&&&&&\\
\cline{1-8}
\cline{10-17}
\cline{19-26}
1&0&&&&&&& &\bf {0}&\bf {0}&&&&&&& &0&1&&&&&&\\
\cline{1-8}
\cline{10-17}
\cline{19-26}
&\bf {0}&\bf {0}&&&&&& &&\bf {0}&\bf {0}&&&&&& &&\bf {0}&\bf {0}&&&&&\\
\cline{1-8}
\cline{10-17}
\cline{19-26}
&&\bf {0}&\bf {0}&&&&& &&1&0&&&&&& &&1&1&&&&&\\
\cline{1-8}
\cline{10-17}
\cline{19-26}
&&&\bf {0}&\bf {0}&&&& &&&\bf {0}&\bf {0}&&&&& &v&&\bf {0}&\bf {0}&&&&\\
\cline{1-8}
\cline{10-17}
\cline{19-26}
&&&&\bf {0}&\bf {0}&&& &&&&\bf {0}&\bf {0}&&&& &w&&1&0&&&&\\
\cline{1-8}
\cline{10-17}
\cline{19-26}
&&&&0&1&&& &&&&0&1&&&& &w&&&\bf {0}&\bf {0}&&&\\
\cline{1-8}
\cline{10-17}
\cline{19-26}
&&&&&\bf {0}&\bf {0}& &&&&&&\bf {0}&\bf {0}& &&&v&&&0&1&&&\\  
\cline{1-8}
\cline{10-17}
\cline{19-26}
&&&&&1&1&& &&&&&1&1&&& &&&&&\bf {0}&\bf {0}&&\\
\cline{1-8}
\cline{10-17}
\cline{19-26}
&&&&&&\bf {0}&\bf {0}& &&&&&&\bf {0}&\bf {0}&& &&&&&1&0&&\\
\cline{1-8}
\cline{10-17}
\cline{19-26}
&&&&&&1&1&   &&&&&&1&0&& &&&&&&\bf {0}&\bf {0}&\\
\cline{1-8}
\cline{10-17}
\cline{19-26}
\multicolumn{8}{}{} &&&&&&&&\bf {0}&\bf {0} &&&&&&&&\bf {0}&\bf {0}\\
\cline{10-17}
\cline{19-26}
\multicolumn{8}{}{}&&&&&&&&&   &&&&&&&&0&1\\
\cline{10-17}
\cline{19-26}                                              

\end{tabular}
\end{table}

\medskip

\bf {Example 3.} \rm The  problem solving for the CCS system containing no JSS (Table 12).

\smallskip

The deduction procedure needs in this case a backtracking for the starting variable $a$ :

 $(a = 0)\Rightarrow (f  = 1) \Rightarrow (g = 0) \Rightarrow (h = 0),$
 
                                 $ (g = 0)\Rightarrow (h = 1)$ --- a contradiction exists for variable $h$;

\medskip
backtracking: 

$(a = 1)\Rightarrow (f  = 0)\Rightarrow (g = 1)\Rightarrow (h = 1),$ 

                                  $(g = 1) \Rightarrow (h = 0)$ --- a    contradiction for variable $h$.
 
Conclusion: no JSS exists.

Owing to simplicity of the deduction the full scheme of  tabular transformations with inverting of the columns and distribution of zeros here is not developed. The logic consequences for two alternative values of variable $a$  are explained in Table 12 by presenting  of the cell pairs, suitable for this purpose, in two different types: thick and italic.

\begin{table}[tbp]
\begin{tabular}{|c|c|c|c|c|c|c|c|c|c|c|c|c|c|c|c|c|c|c|c|c|c|c|c|c|c|}
\multicolumn{26}{c}{\bf Table 12.  CCS system \,$L_1$, $L_2$,  $L_3$}\\
\multicolumn{26}{c}{$L_1$\hspace{5cm}$L_2$\hspace{5cm}$L_3$}\\
\cline{1-8}
\cline{10-17}
\cline{19-26}
$a$&$b$&$c$&$d$&$e$&$f$&$g$&$h$&&$h$&$g$&$b$&$e$&$f$&$c$&$d$&$a$&&
$b$&$c$&$d$&$e$&$a$&$f$&$h$&$g$\\
\hhline{|=|=|=|=|=|=|=|=|~|=|=|=|=|=|=|=|=|~|=|=|=|=|=|=|=|=|}
0&0&&&&&&& &\bf {1}&\bf {1}&&&&&&& &0&1&&&&&&\\
\cline{1-8}
\cline{10-17}
\cline{19-26}
1&0&&&&&&& &\it {0}&\it {0}&&&&&&& &&1&1&&&&&\\
\cline{1-8}
\cline{10-17}
\cline{19-26}
&0&1&&&&&& &&0&0&&&&&& &&&1&1&&&&\\
\cline{1-8}
\cline{10-17}
\cline{19-26}
&&1&1&&&&& &&1&0&&&&&& &&&&1&0&&&\\
\cline{1-8}
\cline{10-17}
\cline{19-26}
&&&1&1&&&& &&&0&1&&&&& &&&&1&1&&&\\
\cline{1-8}
\cline{10-17}
\cline{19-26}
&&&&1&0&&& &&&&1&0&&&& &&&&&\it {0}&\it {1}&&\\
\cline{1-8}
\cline{10-17}
\cline{19-26}
&&&&1&1&&& &&&&1&1&&&& &&&&&\bf {1}&\bf {0}&&\\
\cline{1-8}
\cline{10-17}
\cline{19-26}
&&&&&\bf {0}&\bf {1}& &&&&&&0&1& &&&&&&&&0&1&\\  
\cline{1-8}
\cline{10-17}
\cline{19-26}
&&&&&\it {1}&\it {0}&& &&&&&1&1&& &&&&&&&1&0&\\
\cline{1-8}
\cline{10-17}
\cline{19-26}
&&&&&&1&1& &&&&&&1&1&& &&&&&&&\it {1}&\it {0}\\
\cline{1-8}
\cline{10-17}
\cline{19-26}
&&&&&&0&0&   &&&&&&&1&0&& &&&&&&\bf {0}&\bf {1}\\
\cline{1-8}
\cline{10-17}
\cline{19-26}
\multicolumn{8}{}{} &&&&&&&&1&1 &&&&&&&&&\\
\cline{10-17}
\cline{19-26}
\cline{10-17}
\cline{19-26}                                              

\end{tabular}
\end{table}

\medskip
\textbf{7. The model classificatiion and conclusiions}

\smallskip

Let's address to the formula describing a well-known problem 2-CNF:

\begin{equation}
\bigwedge_{t=1}^{m_2}  (s_{i_t} \vee v_{j_t}),
\end{equation}

where: literals  $s_{i_t}$  and  $v_{j_t}$   designate  different variables of the problem or their inversions, 
$m_2$ is  the number of disjunctions in CNF.

\smallskip
It is difficult not to notice a similarity in the constitution of the formulas (1) and (2); the sign of inversion placed before brackets in (1) is the only formal difference between them.

The formula (2) has been resolved with use of the polynomial algorithm realizing a process of a constraints distribution [4]. The algorithm aims to assign a true ("1") value to each disjunction, consistently choosing suitable values for variables. In more details: the algorithm follows all chains of inference after making each of its choices; this either leads to a contradiction and a backtracking step, or, if no contradiction is derived, it follows that the choice was a correct one that leads to 
a satisfying assignment. Therefore, the algorithm either correctly finds a satisfying assignment or it correctly determines that the input formula is unsatisfiable.

This brief description is in many respects in concordance with the given above description of the procedure of zero values distribution (paying into attention a natural distinction in detailed  operation purposes).

The distinctions in a purpose and realization of steps:
 
•	for the formula (2): to construct at the next step a disjunction with true value taking into account already completed constructions;

•	for the formula (1): to construct at the next step a compact couple $0\,0$ on the basis of system of CCS tables, having generated thereby the next pair of brackets with false values of literals in (1).

Distinctions in the complexity of calculations, following from a data structure, are minimum: the formula (2) is a linear array of the pairs of brackets each containing two literals, coded in CNF; the formula (1) is a two-dimensional array of binary compact couples which may be compared with disjunction in (2).

 So, it is stated that asymptotic computational complexity of both procedures is polynomial. Besides, the same conclusion concerning the procedure of formation of nil-sets in CCS directly follows from Sections 5 and 6.  

\newpage
 As a whole the described algorithmic model of 3-SAT problem representation and solution also belongs to the class of  polynomial models. No use of heuristic modes and complicated systems of hyperstructures radically distinguish the given model from earlier workings out of the author [2, 3], caused a resonance and fairly wide open discussion among programmers.

The paper suggests special constructive components: CTS and CCS, that are unique. Discordance of the structures constructed on the basis of different permutations of variables is overcome by reducing of JSS searching to formation and discerning nil-sets that are common for all structures.

The adaptation of the algorithm of constraints distribution (well-known in the optimization theory) to the efficient resolving of Boolean formula coded by means of discordant compact structures is also a key-point of the investigation.
The results of the work assume a generalization by force of polynomial reducibility among intractable problems. 

\bigskip
\begin{center}
{\bf REFERENCES}
\end{center}

\smallskip 

1. Romanov V. F. Non-orthodox models for the discrete analyses and optimization problems
[Neortodoksal'nyie modeli dlya zadach diskretnogo analiza i optimizatsii]. Saarbr\H {u}cken, Germany, LAP LAMBERT Academic Publishing, 2012. 130 p.

2. Romanov V. F. Non-orthodox combinatorial models based on discordant structures. Electronic journal ``Investigated in Russia", 2007, pp. 1553-1571, available at: 

 http://zhurnal.ape.relarn.ru/articles/2007/143e.pdf

3. Romanov V. F. Non-orthodox combinatorial models based on discordant structures. ArXiv.org., 2011, available at:  http://arxiv.org/abs/1011.3944

4. Lauriere J.-L. Intelligence Artificielle [Sistemy iskusstvennogo intellekta]. Moscow, ``Mir" Publ., 1991. 568 p.

\end{document}